# Sensitive multiplex spectroscopy in the molecular fingerprint 2.4 μm region with a Cr$^{2+}$:ZnSe femtosecond laser

**E. Sorokin [1], I.T. Sorokina [2], J. Mandon [3], G. Guelachvili [3], N. Picqué [3*]**

[1] *Institut für Photonik, TU Wien, Gusshausstrasse 27/387, A-1040 Vienna, Austria, e.sorokin@tuwien.ac.at*
[2] *Department of Physics, NTNU, N-7491 Trondheim, Norway, sorokina@ntnu.no*
[3]*Laboratoire de Photophysique Moléculaire, CNRS, Université Paris-Sud, Bâtiment 350, 91405 Orsay Cedex, France*
* Corresponding author: nathalie.picque@u-psud.fr    http://www.laser-fts.org*

**Abstract:** An ultrashort-pulse Cr:ZnSe laser is a novel broadband source for sensitive high resolution molecular spectroscopy. A 130-fs pulse allows covering of up to 380 cm$^{-1}$ spectral domain around 2.4 μm which is analyzed simultaneously with a 0.12 cm$^{-1}$ (3.6 GHz) resolution by a Fourier-transform spectrometer. Recorded in 13 s, from 70-cm length absorption around 4150 cm$^{-1}$, acetylene and ammonia spectra exhibit a 3800 signal-to-noise ratio and a 2.4·10$^{-7}$ cm$^{-1}$·Hz$^{-1/2}$ noise equivalent absorption coefficient at one second averaging per spectral element, suggesting a 0.2 ppbv detection level for HF molecule. With the widely practiced classical tungsten lamp source instead of the laser, identical spectra would have taken more than one hour.

**OCIS codes:** 140.4050 Mode-locked lasers, 140.3070 Infrared and far-infrared lasers, 140.7090 Ultrafast lasers, 300.6300 Spectroscopy, Fourier transforms, 300.6390 Spectroscopy, molecular, 320.0320 Ultrafast optics

## References and links

## 1. Introduction

Mode-locked lasers and femtosecond frequency combs actively enter [1-6] the field of broadband spectrometry. In particular, their high brightness allows improving the signal to noise ratio (SNR) of the traditional absorption spectroscopy [1]. For spectroscopic applications, the mid-infrared spectral region is of special interest, as most of molecules have intense vibration-rotation transitions lying in this domain. Developing sensitive techniques in this region results therefore in improved detection levels. High resolution is also advantageous because the Doppler profiles of the infrared lines are narrow and their resolved observation increases the sensitivity. The accuracy of the absolute frequency reading is required for unambiguous identification of the lines and species. Additionally, the broadband simultaneous detection is often necessary, for instance, for multi-species trace gas sensing or for accurate renewed molecular theoretical approaches. Most of the traditional spectrometric approaches encounter severe limitations if these four features are to be met simultaneously. For example, the multichannel grating spectrometers have poor efficiency both from the point of view of spectral resolution and broadband coverage; and wavelength agile laser techniques [7] have limited accuracy. In addition, the mid-infrared region has been suffering from a lack of convenient broadband laser sources, while the spectral radiance of incoherent sources is insufficient for sensitive measurements. A broadband mid-infrared approach to high resolution sensitive spectroscopy is therefore highly desirable. A solution to this challenging problem may be found in the multiplex high resolution spectral analysis using the newly developed femtosecond mode-locked lasers and frequency comb sources.

In this paper, we report the implementation of a new spectrometric method based on a femtosecond laser taking the full benefit of the direct access to the mid-infrared wavelength range. We apply the first femtosecond solid-state mode-locked laser source in the molecular fingerprint region above 2 µm to high resolution spectroscopy. The radiation of a $Cr^{2+}$:ZnSe mode-locked laser is absorbed by the gas sample and analyzed by a commercial Fourier transform (FT) spectrometer. Examples of acetylene and ammonia spectra covering up to 380 $cm^{-1}$ in the 4150 $cm^{-1}$ region are shown.

## 2. Experiment

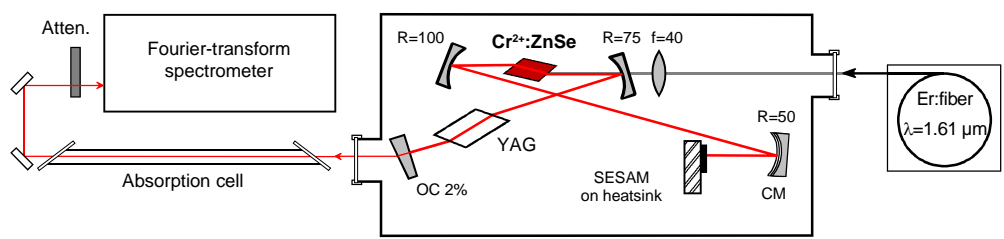

Figure 1: Experimental set-up. CM: chirped mirror. OC: output coupler. Atten: attenuator.

The first sub-ps mode-locked operation of $Cr^{2+}$:ZnSe lasers has been reported recently [8], with pulse duration down to 80 fs at 80 mW of output power [9,10]. In the present experiment, we have set up a prismless femtosecond laser following the design of Ref. [9], optimized for a long-term hands-free stable operation. The laser is based on a 4 mm thick Brewster-cut $Cr^{2+}$:ZnSe crystal (Fig. 1). The crystal temperature is maintained at 20.5 °C by a recirculating water chiller system. The astigmatically compensated X-fold cavity consists of 75-mm and 100-mm radius-of-curvature dichroic mirrors, a 50-mm radius of curvature chirped mirror focusing the light onto an InAs/GaSb semiconductor saturable absorber mirror (SESAM), and an output coupler with transmission of 2 % at 4100 cm$^{-1}$. Optical pumping is achieved with a 1607 nm Er-doped fiber laser, focused onto the crystal by an uncoated lens of 40 mm focal length. The self-starting mode-locking results from the use of the SESAM, which has several hundreds of picoseconds relaxation time. Dispersion compensation is provided by a combination of a 6.5 mm thick uncoated YAG plate and the spherical chirped mirror. This provides about -1500 fs$^2$ round-trip group delay dispersion (GDD), which is almost flat over 500 cm$^{-1}$. The cavity length is about 75 cm, corresponding to 200 MHz pulse repetition rate.

In order to get rid of the strong absorption lines of water vapor, which are present in the vicinity of 2.4 μm, the oscillator is placed inside a sealed enclosure. The enclosure is equipped with windows made of BK7 on the Er:fiber laser side and of CaF$_2$ on the Cr$^{2+}$:ZnSe laser output side. The enclosure is first evacuated under primary vacuum conditions and then filled with dry nitrogen. Under these conditions, with 1.9 W of pumping power, stable mode-locked operation is obtained during hours with ~50 mW average output power. The main stability limitation has been identified as arising from temperature increase of the SESAM heatsink, which was not attached to the cooling circuit, and resulted in slow decrease of the output power (and pulse broadening) during the laser operation. Fig. 2 provides a low resolution (30 GHz - 1 cm$^{-1}$) spectrum and an interferometric autocorrelation trace of the laser pulses. The full width at half maximum (FWHM) of the autocorrelation trace is 250 fs, which leads to a pulse width of the order of 130 fs, consistent with the 76 cm$^{-1}$ measured FWHM of the corresponding spectrum, which has almost perfect sech$^2$ profile, as shown on Fig. 2. The pulse duration could be decreased by dispersion optimization [9], but we opted to improve the operation stability by providing some additional negative GDD.

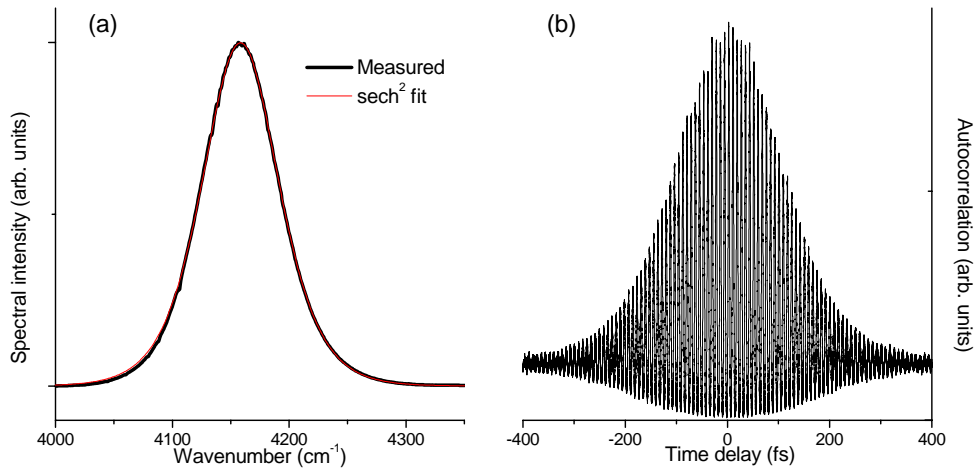

Figure 2: (a) Spectrum (FWHM: 76 cm$^{-1}$) and (b) interferometric autocorrelation (FWHM: 250 fs) of a 130-fs pulse of the Cr$^{2+}$:ZnSe mode-locked laser. Pump power: 1.9 W, output power: 50 mW, repetition rate 200 MHz.

The laser emission is centered at 4157 cm$^{-1}$. As compared to the open-air operation, the emission peak is not shifted towards longer wavelengths. This is different from our previous observation of the non-stationary continuous-wave operation of the Cr$^{2+}$:ZnSe laser under secondary vacuum [11], where the decrease of the water absorption (growing towards smaller wavenumbers) after air removal resulted in a shift of the laser wavelength by ~75 cm$^{-1}$ to the infrared. More surprisingly, the pulse spectral width also doesn't increase when the air humidity absorption disappears. Our explanation is that the high losses of the mode-locked setup (12% unsaturated absorption in the SESAM alone) significantly exceed the integrated absorption in the narrowband water vapor lines, thus making the setup almost insensitive to the air humidity.

After the exit window, the laser pulses pass through a 70 cm-long single-pass absorption cell, filled with the gas of interest, and ~5 m of open-air propagation before finally reaching the spectrometer. The laser radiation is analyzed by a commercial FT spectrometer (max. resolution 0.12 cm$^{-1}$) equipped with a fluorine beam-splitter and a thermoelectrically cooled InAs detector. The laser beam has to be attenuated in order not to saturate the detector.

As a first demonstration of the capabilities of the source for broadband absorption spectroscopy, the rovibrational spectra of acetylene and ammonia have been recorded around 4150 cm$^{-1}$. Figure 3 gives an illustration of the $C_2H_2$ spectrum, in the region of the $\nu_1 + \nu_5^1$ band [12]. The full width at half maximum of the spectrum is 75 cm$^{-1}$. Resolution is limited by the spectrometer to 3.6 GHz (0.12 cm$^{-1}$). With the total recording time $T=13$ s we benefit from an uppermost SNR better than 3800. With spectral boundaries defined by a SNR relatively degraded by a factor 1000, the spectral range extends over 380 cm$^{-1}$, from 3980 cm$^{-1}$ to 4360 cm$^{-1}$. This corresponds to a number of spectral elements $M$ equal to 3167. The cell is filled with acetylene at 23 hPa (17 Torr) pressure. The absorption lines shown on Fig. 3 are due to the acetylene bands and the water vapor lines [13] from the open-air propagation. Figure 4 displays a portion of a $NH_3$ recorded absorption spectrum, with a $NH_3$ pressure equal to 261 hPa (196 Torr). The spectral lines of ammonia belong to the $\nu_1 + \nu_2$ and $\nu_2 + \nu_3$ rovibrational combination bands [14]. As the spectrum is complex and crowded, high resolution is essential to discriminate against the various lines. The recording conditions are the same as for the acetylene spectrum.

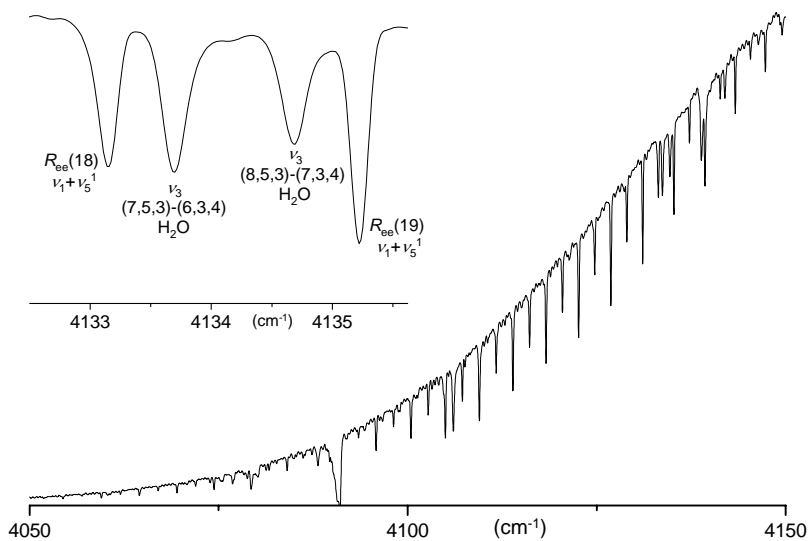

Figure 3: Part of a $C_2H_2$ spectrum at 0.12 cm$^{-1}$ resolution in the 2.4 µm region. The baseline modulation is not noise-related but due to a channel spectrum from an unwedged window. The strongest spectral features are the $P$, $Q$ and $R$ branches of the $\nu_1 + \nu_5^1$ band [12] of $^{12}C_2H_2$, centered at 4090 cm$^{-1}$. The intensity alternation is obvious in the $R$-branch. The upper left inset is a zoom showing acetylene and residual water vapor lines coming from a 5 meters air path between the cell and the interferometer. The $H_2O$ lineshapes reveal an atmospheric-pressure broadening. The rotational assignment for the $H_2O$ lines [13] gives $(J, K_a, K_c)$.

## 3. Results and discussion

In these spectra, the noise equivalent absorption coefficient (NEA) reaches $3.7 \times 10^{-6}$ cm$^{-1}$. NEA at one second averaging per spectral element $(L \times \text{SNR})^{-1} \times (T/M)^{1/2} = 2.4 \times 10^{-7}$ cm$^{-1}$. Hz$^{-1/2}$, where $L$ is the absorption path length. In the present spectral region, this level of sensitivity allows for the high resolution detection of 22 parts per billion by volume (ppbv) of $C_2H_2$, and 160 ppbv of $NH_3$ at 1 s of integration time per spectral element. These detection limit values are still relatively high because the combination bands of $C_2H_2$ and $NH_3$ in this region are weak: their intensity, of the order of $10^{-21}$ cm·molecule$^{-1}$, is not higher than in the 1.5 μm region. For hydrogen fluoride HF (1-0 band, with intensity of the R(2) line at 4075 cm$^{-1}$ equal to $2.3 \cdot 10^{-18}$ cm·molecule$^{-1}$ [13]), the corresponding detection level is 200 parts per trillion by volume (pptv) in the 2.4 μm spectral region.

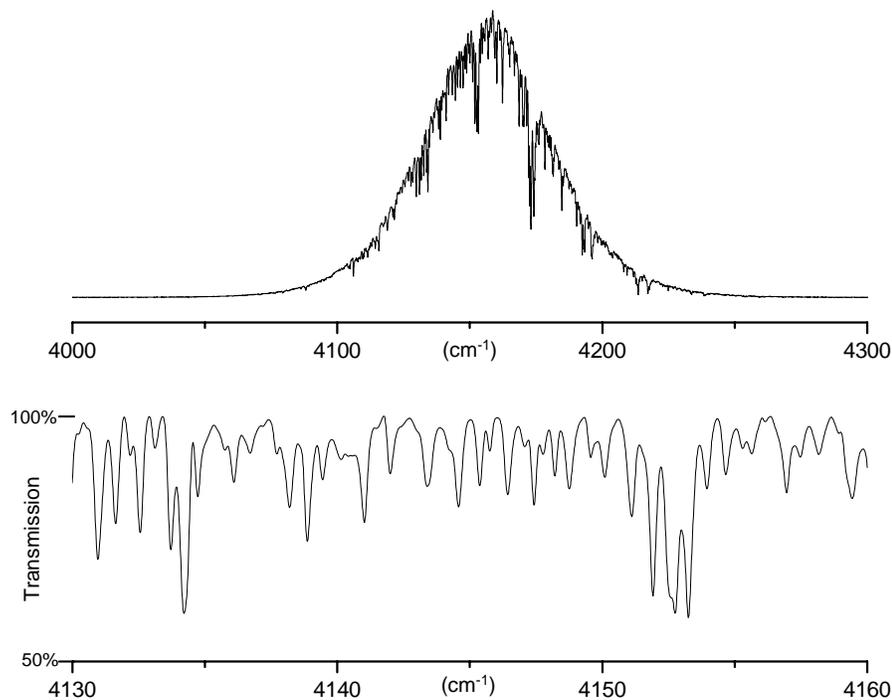

Figure 4: Spectrum of $NH_3$ in the 2.4 μm region illustrating the spectral bandwidth and SNR capabilities of the spectrometric technique. The $\nu_1 + \nu_2$ and $\nu_2 + \nu_3$ combination bands are observed. The upper part of the figure shows the whole spectral domain covered in a singe recording while the lower part shows a portion expanded in wavenumber and intensity scales.

For a fair comparison, we recorded a spectrum using a traditional tungsten lamp, which is the most widely used broadband source in the near- and mid-infrared absorption FTS. Under identical experimental conditions, the SNR was degraded by about 17. The recording time would have then been 300 times longer to get identical results. This is not surprising as this 50 mW femtosecond laser source has a spectral radiance which is about $2.8 \cdot 10^5$ times stronger than a 3000 K blackbody source. Actually, a signal-to-noise ratio enhancement of about 500 should be achieved. This could not be demonstrated in the present experiment: as the 50 mW power of the laser saturates the detector, we had to attenuate the beam by more than two orders of magnitude before entering in the interferometer.  To make a proper use of the abundant power, one should e.g. employ a classical multipass cell. In this case increase of the absorption path by more than two orders of magnitude would then improve the sensitivity by the same factor, without lowering the SNR.

It is also interesting to compare our experimental results with a recent near-infrared comb-based experiment [2]. By coupling a 1550 nm Er-fiber based frequency comb to a cavity with a finesse higher than 3100, Ref. [2] obtains a NEA at one second averaging equal to $2 \cdot 10^{-8}$ cm$^{-1}$.Hz$^{-1/2}$, with an equivalent absorption path length higher than 3000 m. This absorption path length is about 4300 times higher than ours whereas the resulting sensitivity is only one order of magnitude better. Moreover, the resolution of the grating spectrometer is limited to 25 GHz. The acquisition of a broad spectral domain is a sequential, long and inaccurate process, whereas in the present experiment, 13 seconds are enough to measure simultaneously 3167 individual spectral elements spanning 380 cm$^{-1}$, that are only limited by the laser emission, at 3.6 GHz resolution. Actually, during these 13 seconds, the FT spectrometer samples a 15800 cm$^{-1}$ broad spectral domain by 131650 independent spectral elements. A more detailed discussion of the advantages of using a FT spectrometer for this kind of experiments may be found in [1].

Further improvements of the present experiment can be made in several directions. The resolution can easily be improved up to the Doppler width of the lines by using a higher resolution interferometer. It is also possible to significantly increase the measured spectral domain. The enormous bandwidth of the Cr$^{2+}$:ZnSe laser allows indeed coverage of a much broader spectral region, either by tuning, or more interestingly by construction of a few-cycle laser source to observe hundreds of nm simultaneously. Alternatively generation of a mid-infrared supercontinuum in highly nonlinear fibers looks promising, as we demonstrated it recently in the near infrared spectral region [15]. Using of a broadband source is especially attractive for FT spectroscopy, where measurement time is independent of the spectral domain width. As discussed above, sensitivity improvement can benefit from absorption path length enhancement either by use of a classical multipass cell or by injection of a fs frequency comb in a high finesse cavity. Use of a comb can bring additional sensitivity gain by performing high frequency synchronous detection, which also leads to the advantage of allowing simultaneous measurement of the absorption and dispersion associated with the spectral features, as experimentally established in [6].

## 4. Conclusion

Summarizing, we have demonstrated the first spectroscopic application of a mid-infrared femtosecond mode-locked laser used as a broadband infrared source with a high resolution FT spectrometer. This simple experimental setup already exhibits high sensitivity and resolution over a broad spectral domain. Sub-ppb detection levels should be easily obtained for a large panel of molecules thanks to absorption path length enhancement and high frequency detection. Further improvements in resolution and acquisition time are also in progress. This will include taking advantage of the comb structure of a stabilized mode-locked Cr$^{2+}$:ZnSe laser by combination of the complementary methods developed in [4] and in [6].

**Acknowledgments**

P. Jacquet (Laboratoire de Photophysique Moléculaire, Orsay) is warmly thanked for his participation to the experiments. I. T. Sorokina acknowledges Université Paris-Sud for a position of invited professor. This work is accomplished in the framework of the Programme Pluri-Formation de l'Université Paris-Sud "Détection de traces de gaz" 2006-2009 and the Austrian Fonds zur Förderung der wissenschaftlichen Forschung project P17973.